\documentclass[sigconf]{acmart}
\AtBeginDocument{%
  \providecommand\BibTeX{{%
    \normalfont B\kern-0.5em{\scshape i\kern-0.25em b}\kern-0.8em\TeX}}}

\setcopyright{acmcopyright}
\copyrightyear{2023}
\acmYear{2023}
\acmDOI{XXXXXXX.XXXXXXX}

\acmConference[Conference acronym 'XX]{Make sure to enter the correct
  conference title from your rights confirmation emai}{June 03--05,
  2018}{Woodstock, NY}
\acmPrice{15.00}
\acmISBN{978-1-4503-XXXX-X/18/06}

\usepackage{hyperref}
\usepackage{cleveref}
\usepackage{subcaption}
\usepackage{enumitem}
\usepackage{algorithm}
\usepackage{algpseudocode}
\usepackage{amsmath}
\usepackage{bbm}
\begin{document}

\title{RankTower: A Synergistic Framework for Enhancing Two-Tower Pre-Ranking Model}

\author{YaChen Yan}
\email{yachen.yan@creditkarma.com}
\orcid{0000-0002-1213-4343}
\affiliation{%
    \institution{Credit Karma}
    \streetaddress{760 Market Street}
    \city{San Francisco}
    \state{California}
    \country{USA}
    \postcode{94012}
}

\author{Liubo Li}
\email{liubo.li@creditkarma.com}
\orcid{1234-5678-9012}
\affiliation{%
    \institution{Credit Karma}
    \streetaddress{760 Market Street}
    \city{San Francisco}
    \state{California}
    \country{USA}
    \postcode{94012}
}

\renewcommand{\shortauthors}{YaChen and LiuBo}
\renewcommand{\subtitle}{RankTower}

\begin{abstract}
In large-scale ranking systems, cascading architectures have been widely adopted to achieve a balance between efficiency and effectiveness. The pre-ranking module plays a vital role in selecting a subset of candidates for the subsequent ranking module. It is crucial for the pre-ranking model to maintain a balance between efficiency and accuracy to adhere to online latency constraints. In this paper, we propose a novel neural network architecture called RankTower, which is designed to efficiently capture user-item interactions while following the user-item decoupling paradigm to ensure online inference efficiency. The proposed approach employs a hybrid training objective that learns from samples obtained from the full stage of the cascade ranking system, optimizing different objectives for varying sample spaces. This strategy aims to enhance the pre-ranking model's ranking capability and improvement alignment with the existing cascade ranking system. Experimental results conducted on public datasets demonstrate that RankTower significantly outperforms state-of-the-art pre-ranking models.
\end{abstract}

\ccsdesc[500]{Computing methodologies}
\ccsdesc[300]{Machine learning}
\ccsdesc{Machine learning approaches}
\ccsdesc[100]{Neural networks}

\keywords{Recommender Systems, Pre-Ranking, Learning to Rank, Differentiable Sorting}

\maketitle

\section{Introduction}

\begin{figure}
    \centering
    \includegraphics[width=0.5\textwidth, height=0.1\textwidth]{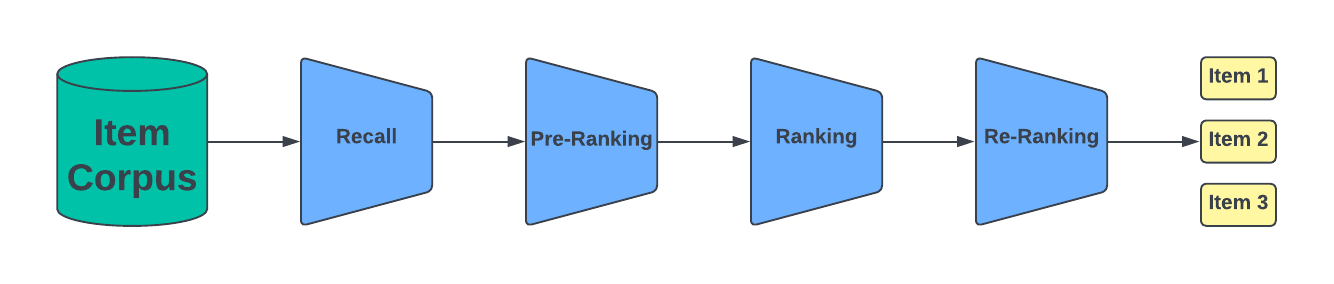}
    \caption{The Architecture of Cascade Ranking System}
    \label{fig_ranking_flow}
\end{figure}

In industrial information services, such as recommender systems, search engines, and advertisement systems, the cascading architecture ranking system has been widely used to achieve a balance between efficiency and effectiveness. A typical cascade ranking system, as illustrated in \autoref{fig_ranking_flow}, consists of multiple sequential stages, including recall, pre-ranking, ranking, and re-ranking stages.

Pre-ranking is commonly regarded as a lightweight ranking module characterized by a simpler network architecture and a reduced set of features. Compared to ranking models, pre-ranking models are required to score a larger number of candidate items for each user and demonstrate higher inference efficiency, although their prediction performance may be comparatively weaker due to their simpler structure. Given the emphasis on efficiency, pre-ranking typically employs a straightforward vector-product-based model.

We propose a novel framework for pre-ranking systems to maintain consistency with the cascade ranking system and achieve a balance between inference efficiency and prediction accuracy. The primary contributions of this work are as follows:

\begin{itemize}
    \item We introduce the RankTower architecture, which comprises three key components: Multi-Head Gated Network, Gated Cross-Attention Network, and Maximum Similarity Layer. The Multi-Head Gated Network plays a vital role in extracting diversified latent representations of users and items. The Gated Cross-Attention Network enables the modeling of bi-directional user-item interactions. Finally, the Maximum Similarity Layer enhances online serving efficiency without compromising the model's performance.
    \item We employ a full-stage sampling strategy by drawing the training samples from different stages of the cascade ranking system. Tightly coupled with this sampling approach, we strategically integrate a hybrid loss function that judiciously combines distillation and learning-to-rank losses. This synergistic approach facilitates comprehensive learning of the ordering dynamics underlying user interactions while accounting for the inherent characteristics of the cascade ranking system.
    \item We conduct extensive experiments on three publicly available datasets to demonstrate the superior performance of RankTower in terms of prediction accuracy and inference efficiency.
\end{itemize}

\section{Related Work}

In this section, we provide an overview of the most recent studies on pre-ranking model, which serve as a crucial intermediary stage in the cascading ranking system. The primary function of the pre-ranking stage is to effectively reduce the large pool of candidates retrieved from the recall stage to a manageable subset for the subsequent ranking stage. Furthermore, we discuss existing point-wise, pair-wise, and listwise ranking losses commonly employed in training learning-to-rank models.

\subsection{Pre-Ranking}
Several studies propose to improve the efficiency and accuracy of the pre-rank system. COLD \cite{wang2020cold} is designed to jointly optimize both the pre-ranking model performance and the computing power it costs. Any arbitrary deep model with cross features can be applied in COLD under a constraint of controllable computing power cost. Computing power cost can also be explicitly reduced by applying optimization tricks for inference acceleration. FSCD \cite{ma2021towards} achieves a better trade-off between effectiveness and efficiency by utilizing the learnable feature selection method based on feature complexity and variational dropout. AutoFAS \cite{li2022autofas} selects the most important features and network architectures using Neural Architecture Search, and a ranking model guided reward is equipped during NAS procedure, which allows AutoFAS to select the best pre-ranking architecture for a given ranking teacher without any computation overhead. IntTower \cite{li2022inttower} is designed to address the efficiency-accuracy dilemma in pre-ranking systems. The proposed IntTower achieves high prediction accuracy while maintaining inference efficiency by balancing the interactions between user and item representations.

Another direction of research is to align the pre-ranking with the ranking prediction order and ranking stages. RankFlow \cite{qin2022rankflow} and Ranking Distillation \cite{tang2018ranking} have proposed aligning the pre-ranking and ranking models through distillation based on ranking scores. The pre-ranking model is encouraged either to generate the same scores as the ranking model\cite{qin2022rankflow} or to produce high scores for the top candidates selected by the ranking model\cite{tang2018ranking}. JRC \cite{sheng2022joint} introduces an approach called Jointly Ranked Calibration that optimizes both ranking and calibration abilities. JRC enhances the ranking ability by comparing the logit values for a sample with different labels and ensures the predicted probability is constrained as a function of the logit subtraction. COPR \cite{zhao2023copr} optimizes the pre-ranking model towards consistency with the ranking model. It employs a chunk-based sampling module and a plug-and-play rank alignment module to explicitly optimize the consistency of ECPM-ranked results. Recently, \cite{wang2023adaptive} employed relaxed sorting loss to directly maximize business goals on ranking stage level.

\subsection{Learning-to-Rank Losses}
Learning-to-Rank(LTR) losses are typically categorized into three main types: pointwise, pairwise, and listwise. Each category reflects a different approach to how the ranking problem is formulated and optimized. \cite{li2011short, liu2009learning}

The pointwise approach treats the ranking problem as a classification or regression task, aiming to predict the relevance score of each item independently without considering the relative order among items or the user-item group structure. The typical industry solutions employ binary logistic loss due to the binary nature of most user feedback, and \cite{covington2016deep} adapted this approach for regression tasks.

The pairwise approach transformed the ranking problem into pairwise classification or pairwise regression. It focuses on optimizing the relative order of item pairs but still overlooks the user-item group structure. BPR loss \cite{rendle2012bpr} utilizes binary logistic loss to model the probability that one item is ranked higher than another. WARP loss \cite{weston2011wsabie} further incorporates rank-based weighting to prioritize the accurate ranking of the most relevant items. LambdaRank \cite{burges2010ranknet} extends RankNet \cite{burges2005learning} by re-weighting the gradients of the loss function based on the impact of changes in NDCG metrics.

The listwise approach directly optimize the ranking problem
on user-item group structure. ListMLE \cite{xia2008listwise} optimizes the likelihood of the correct permutation based on the predictions. ListNet \cite{cao2007learning} utilizes softmax cross entropy to learn the probability distribution over permutations. \cite{gupta2023efficacy} further proposed decoupled softmax loss to address limitations in traditional softmax loss for extreme multi-label scenario. ApproxNDCG \cite{qin2010general, bruch2019revisiting} optimizes NDCG metric with a differentiable approximation based on the logistic function. NeuralSort \cite{grover2019stochastic} and SoftSort \cite{prillo2020softsort}, initially designed for differentiable sorting, were later adapted to solve ranking problems.

\section{Model Architecture}

\begin{figure}
    \centering
    \includegraphics[width=0.45\textwidth, height=0.45\textwidth]{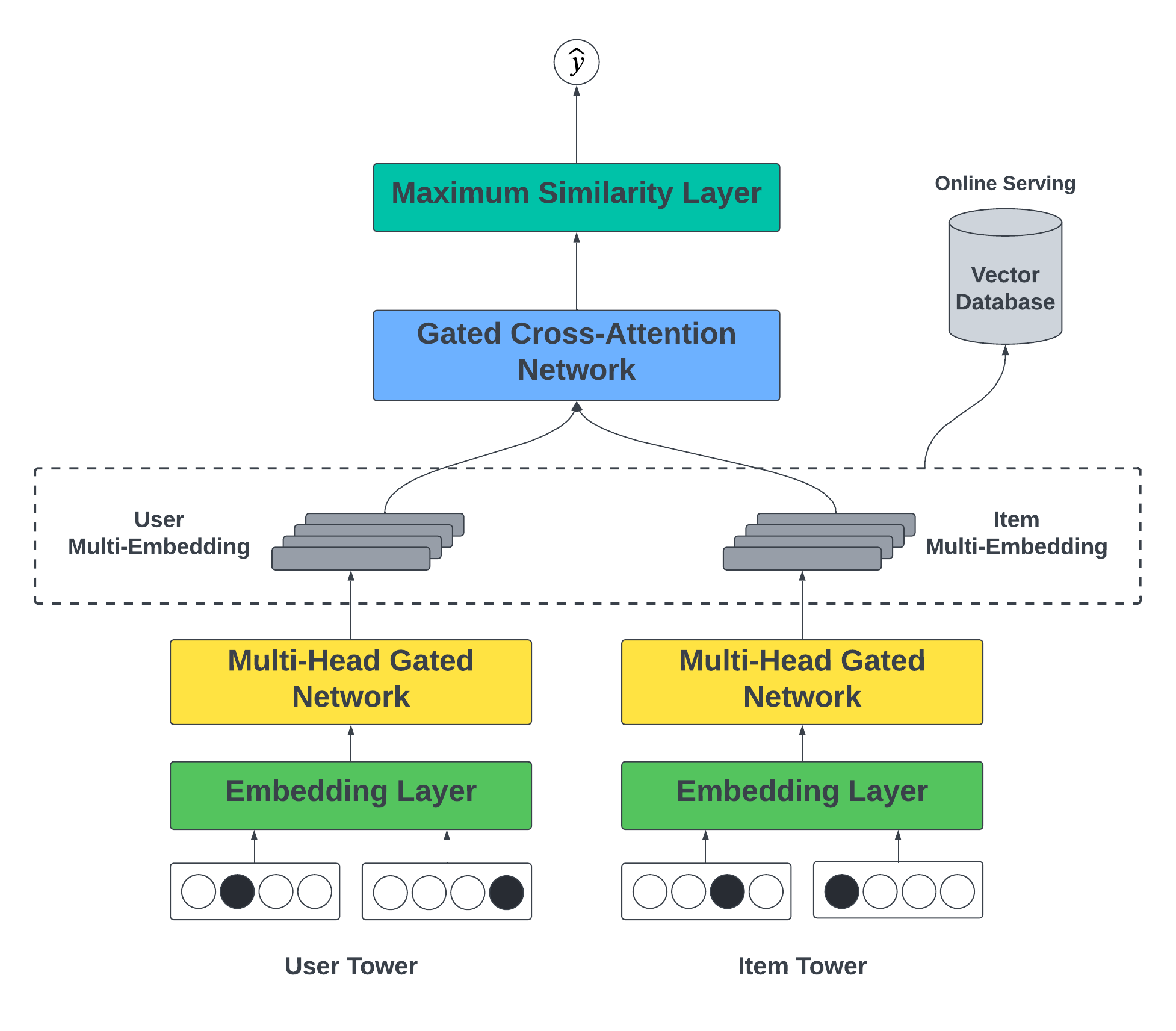}
    \caption{The Architecture of RankTower}
    \label{fig_ranktower}
    \medskip
    \small
\end{figure}

The RankTower architecture, shown in  \autoref{fig_ranktower}, introduces three main modules besides the embedding layer: Multi-Head Gated Network, Gated Cross-Attention Network, and Maximum Similarity Layer. The Multi-Head Gated Network computes diversified user and item representations by dynamically identifying feature importance. The Gated Cross-Attention Network models bi-directional user-item interactions, and the Maximum Similarity Layer efficiently captures the interaction between user-attentive and item-attentive embeddings to compute the final prediction.

RankTower follows the user-item decoupling paradigm, enabling efficient online serving by pre-computing and storing user and item representations into a vector database. During online serving, only the gated cross-attention layers require forward propagation, while other operations remain parameter-free, optimizing computational efficiency.

\subsection{Preliminary}
The dataset for building the pre-ranking model consists of instances $(x_{u}, x_{i}, y, p)$, where $x_{u}$ and $x_{i}$ represent the user feature and item feature respectively, $y \in \{0, 1\}$ indicates the user-item binary feedback label, $p$ is the logged ranking model probability prediction that will be used for training the pre-ranking model with knowledge distillation. $z$ is the logit of the pre-ranking model and $\hat{y}$ is the corresponding pre-ranking prediction.

\subsection{Embedding Layer}
Suppose we have $F_{U}$ fields of user features and $F_{I}$ fields of item features in our pre-ranking training data. In our feature processing step, we first bucketize all the continuous features to equal frequency bins, then embed the bucketized continuous features and categorical features embed each feature onto a dense embedding vector $x$. Lastly, we concatenate $F_{U}$ and $F_{I}$ embedding vectors separately and denote the output of embedding layer $X_{U}$ and $X_{I}$ as the user input embedding and item input embedding, respectively:

\begin{align}
    X_{U} &= [\mathbf{x}_{u}^{1}, \mathbf{x}_{u}^{2}, \cdots, \mathbf{x}_{u}^{F_{U}}]^T. \\
    X_{I} &= [\mathbf{x}_{i}^{1}, \mathbf{x}_{i}^{2}, \cdots, \mathbf{x}_{i}^{F_{I}}]^T.
\end{align}

\subsection{Multi-Head Gated Network}


The Multi-Head Gated Network is an improved MLP augmented with a gating mechanism, mainly for extracting diversified user and item representations from user/item input embeddings.

We first use MLP to model the deep user/item representations and further multiply the output of the MLP by an instance-aware gating vector. The gating vector can be modeled by a two-layer MLP with a reduction ratio $r$, using user/item input embeddings as input. During the training phase, the input embedding does not receive gradients from the gating network to ensure training stability. 

For example, given an user input embedding $X_{U}$, the output of user multi-embedding can be mapped into $H_{u}$ sub-spaces, and the $h$-th sub-spaces $e_{u}^{h}$ is obtained from:

\begin{equation}
\begin{aligned}
    e_{u}^{h} & = MLP_{u}(X_{U})^{h} \circ \sigma(gMLP_{u}(X_{U}))^{h} \\
    & \in \mathbb{R}^{B \times k}, \qquad h=1,\cdots, H_{u}
\end{aligned}
\end{equation}

where $\circ$ denotes the Hadamard product, $\sigma$ denotes the activation function of the gating network: $\mathrm{Sigmoid}(x)$, $MLP_{u}$ denotes the MLP layer for modeling the user input embedding and extracting the latent information, $gMLP_{u}$ denotes the gating MLP for facilitating selective attention, $B$ is the batch size and $k$ is the embedding size of each sub-space.

Similarly, given an user input embedding $X_{I}$, the output of item multi-embedding can be mapped into $H_{i}$ sub-spaces, and the $h$-th sub-spaces $e_{i}^{h}$ is obtained from:

\begin{equation}
\begin{aligned}
    e_{i}^{h} & = MLP_{i}(X_{I})^{h} \circ \sigma(gMLP_{i}(X_{I}))^{h} \\
    & \in \mathbb{R}^{B \times k}, \qquad h=1,\cdots, H_{i}
\end{aligned}
\end{equation}

In the offline processing stage, we will periodically batch inference and store all the user/item's embeddings $e_{u}^{h}$ and $e_{i}^{h}$ into the vector database for online serving usage.

\subsection{Gated Cross-Attention Network}
The Gated Cross-Attention Network employs the cross-attention mechanism to effectively model the interaction between user embedding and item embedding. It uses the Gated Attention Unit as the main building block for learning the dependency between user and item, with residual connection and layer normalization used for training stability.

\subsubsection{Cross Attention Mechanism}

\begin{figure}
    \centering
    \includegraphics[width=0.40\textwidth, height=0.40\textwidth]{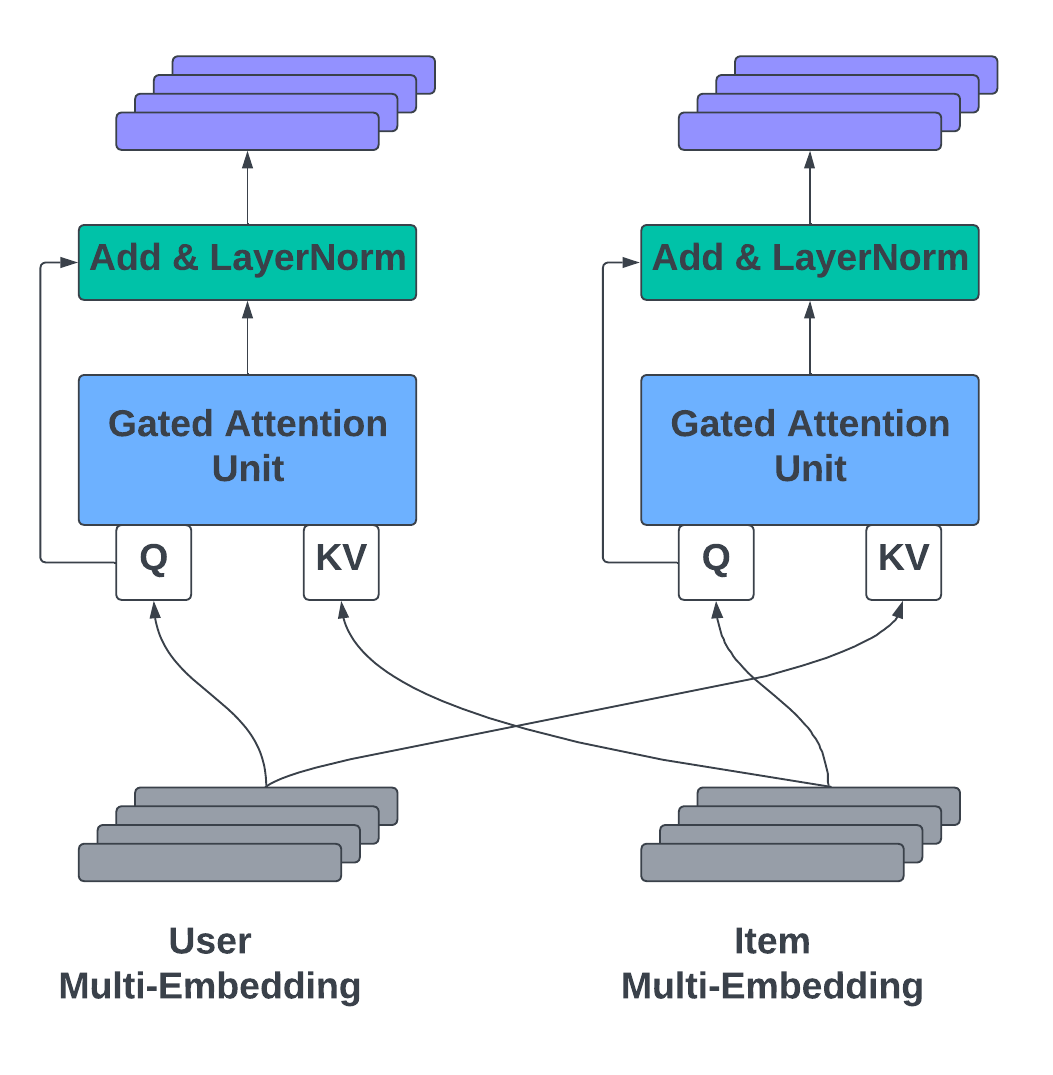}
    \caption{The Architecture of Gated Cross-Attention Network}
    \label{fig_cross_attention}
    \medskip
    \small
\end{figure}

The Bi-Directional Gated Cross-Attention Network interchangeably uses user and item embedding as queries and keys-values for bi-directional attention. Specifically, with the user multi-embedding $E_{u} = \mathrm{Concat}(\mathrm{e_{u}^{1}}, ..., {e_{u}^{H_{u}}})$ and item multi-embedding $E_{i} = \mathrm{Concat}(\mathrm{e_{i}^{1}}, ..., {e_{i}^{H_{i}}})$, the cross-attention compute the user attended embedding $\mathcal{E}_{u}$ and item attended embedding $\mathcal{E}_{i}$ as follows:

\begin{align}
    \mathcal{E}_{u} &= \mathrm{LayerNorm}(E_{u} + \mathrm{GAU}(Q=E_{u}, K=E_{i}, V=E_{i})) \in \mathbb{R}^{B \times H_{u} \times k} \\
    \mathcal{E}_{i} &= \mathrm{LayerNorm}(E_{i} + \mathrm{GAU}(Q=E_{i}, K=E_{u}, V=E_{u})) \in \mathbb{R}^{B \times H_{i} \times k}
\end{align}

The cross-attention mechanism with two parallel branches is designed to process information from both user embedding and item embedding simultaneously. By having two parallel branches, one focusing on user information and the other on item information, the model can simultaneously attend to both user preferences and item characteristics. This enables the model to capture bidirectional interactions between user embedding and item embedding, leading to more accurate modeling of user-item interactions. The overall structure is shown in \autoref{fig_cross_attention}.

\subsubsection{Gated Attention Unit}


The Gated Attention Unit introduces a gating mechanism to facilitate selective attention for better learning the dependency between user embedding and item embedding. Specifically, the Gated Attention Unit effectively enables an attentive gating mechanism as follows:

\begin{align}
    Q &= \phi(X_{Q} W_{Q}) \\
    K &= \phi(X_{K} W_{K}) \\
    V &= \phi(X_{V} W_{V}) \\
    U &= \sigma(X_{Q} W_{U})
\end{align}

where $X_{Q}$, $X_{K}$, $X_{V}$ are the query, key, and value input, $\phi$ is the non-linear activation function for query, key and value projection, $\sigma$ is the sigmoid function for computing gating value based on the query.

With the learned projection $Q$, $K$, $V$, and the gating value $U$, we compute the attention weights, followed by gating and a post-attention projection.

\begin{align}
    O &= (U \odot AV)W_{o} \\
    A &= \mathrm{softmax}(\frac{QK^T}{\sqrt{d_k}})
\end{align}

where $A \in \mathbb{R}^{H_{u} \times H_{i}}$ contains user to item attention weights. This example assumes that we use user embedding as the query, and item embedding as key and value.

\subsection{Maximum Similarity Layer}
The Maximum Similarity Layer computes the final probability prediction based on the user attended embedding $\mathcal{E}_{u}$ and item attended embedding $\mathcal{E}_{i}$. Specifically, each user sub-space firstly computes maximum cosine similarity with all the item sub-space; the scalar outputs of these operations are summed across all the user sub-spaces:

\begin{equation}
\begin{aligned}
    s = (\sum_{p=1}^{H_{u}} \underset{q\in \{1,\cdots, H_{i}\}}{\mathrm{Max}} COSINE(\mathcal{E}_{u}^{p}, \mathcal{E}_{i}^{q})) / \tau
\end{aligned}
\end{equation}

where $p$ and $q$ are the sub-space indexes of user-attended embedding and item-attended embedding, respectively, and $\tau$ is the learnable temperature scalar for re-scaling the cosine similarity.

Note that the Maximum Similarity Layer does not have any parameters which is suitable for online serving.

\section{Pre-Ranking Model Optimization}

\begin{figure*}
    \centering
    \includegraphics[width=0.80\textwidth, height=0.40\textwidth]{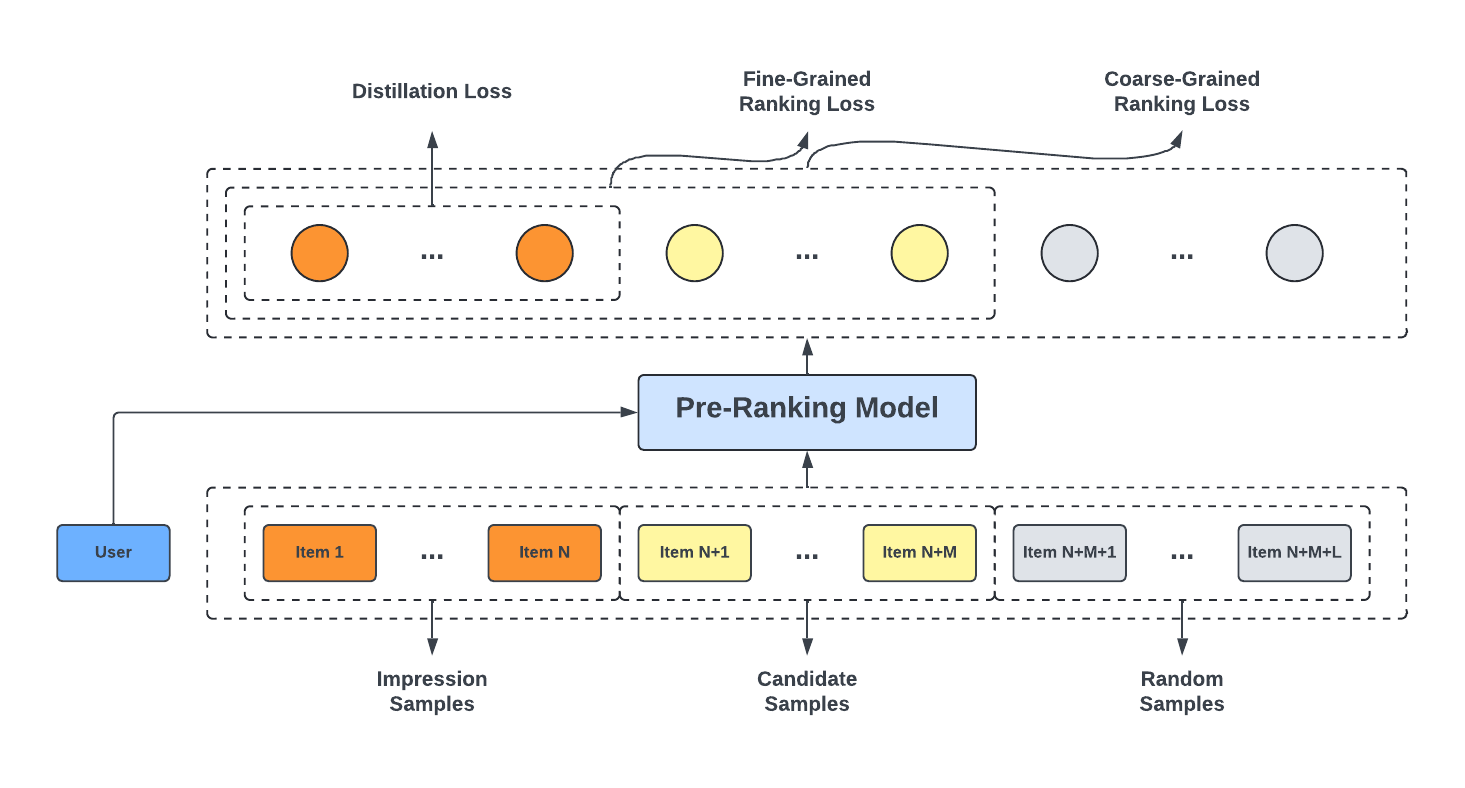}
    \caption{The Synergistic Framework for Learning User Behavior Ordering and Full-Stage Sample Ordering}
    \label{fig_sample_loss}
    \medskip
    \small
\end{figure*}

Despite the significance of enhancing the consistency between ranking models and pre-ranking models, pre-ranking models trained exclusively on impression samples, same as ranking models, suffer from sample selection bias. The pre-ranking model, which operates on the outputs of recall models, aims to identify the most relevant candidates set for the ranking model. Consequently, aligning the item distribution between the training and serving phases is essential to mitigate this sample selection bias and improve model effectiveness.

As illustrated in \autoref{fig_sample_loss}, we implemented full-stage sampling to draw training data from impression samples, candidate samples, and random samples to mitigate sample selection bias. Moreover, we strategically applied various distillation and learning-to-rank losses to different sample scopes to effectively learn the ordering of user behaviors and the sequencing of the sample stages.

\subsection{Full-Stage Sampling}
The RankTower model is trained using user-level listwise samples that comprise multiple positive items and multiple objectives. The training samples associated with each user are sourced from various stages of the cascade ranking system, as shown in \autoref{fig_ranking_flow}. Detailed definitions and relationships among these components are provided below:

\subsubsection{Impression Samples}
The items output by the ranking model and viewed by the user consist of both positive and negative samples. Positive samples are items that have received various types of positive user feedback. In contrast, negative samples are items that have been exposed to the user but have not received any positive user feedback.

\subsubsection{Candidate Samples}
The item candidates in the ranking or pre-ranking stages that are not viewed by the user are categorized based on their progression through the cascade ranking pipeline. Ranking candidates, which have advanced to the ranking stage, are generally considered as hard negative samples due to their higher relevance and quality compared to the pre-ranking candidates. In contrast, pre-ranking candidates are regarded as relatively easy negative samples because they were filtered out before reaching the ranking stage, indicating a lower level of relevance or potential interest to the user.

\subsubsection{Random Samples}
Items that are randomly sampled from the item corpus to serve as negative samples. These random samples are considered the easiest negative samples but are included to further enhance the generalization capability of the pre-ranking model. The incorporation of random samples ensures that the model remains effective and adaptable when encountering previously unseen items during the serving phase, thereby improving its robustness and ability to handle diverse item distributions.

\subsection{Label Aggregation}
Our framework incorporates two categories of labels: hard labels and soft labels. Hard labels represent various types of positive user feedback on impression samples. Soft labels are the predictions made by the ranking model for different types of user feedback. These soft labels are utilized for knowledge distillation to improve the consistency between pre-ranking models and ranking models, ensuring that the pre-ranking model learns to mimic the behavior of the ranking model. Both categories of labels require an aggregation function to consolidate the different user behaviors into a single scalar value for the pre-ranking model's learning.

\subsubsection{Hard Labels}
The aggregation of hard labels is highly dependent on the specific business problem, requiring that labels be aggregated according to their orders of importance.

For instance, in online advertising, eCPM can be utilized based on the pricing model of the platform. In an e-commerce context, one might establish a relative preference order based on the depth of user feedback, such as \textit{Purchase} > \textit{Add to Cart} > \textit{Click}. For scenarios like feed ranking or video recommendations, users may provide various types of feedback signals. These signals can be aggregated using a weighted sum approach, based on the specific business objectives. In addition to user feedback labels, we incorporate a general impression label applicable across business scenarios, for learning the pattern of the cascade ranking system. The label assigned a value of 1 for impression samples and 0 otherwise. It is important to note that all the candidate samples and random samples are considered negative in terms of hard labels.

The user feedback labels help the pre-ranking model in learning the revenue or engagement level associated with different user behaviors. The exposure label facilitates the pre-ranking model's ability to learn and replicate the ranking patterns in the downstream cascade ranking system.

\subsubsection{Soft Labels}
Similar to the aggregation of hard labels, the aggregation of soft labels should employ the ranking objective function. This approach ensures that the soft labels, which are derived from the predictions of the ranking model, are seamlessly integrated into the training process. By utilizing the ranking objective function, the consistency between the pre-ranking model and the ranking model is maintained.

\subsection{Hybrid Loss Functions}
We design the pre-ranking model learning strategy to achieve two primary objectives: consistency and ranking.

To guide and accelerate the model training process, we utilize the ranking model's predictions as soft labels for the purpose of knowledge distillation from the ranking model (serving as the teacher model) to the pre-ranking model (serving as the student model). This approach ensures consistency between the pre-ranking and ranking stages.

To facilitate both ranking accuracy and retrieval capability, we apply fine-grained and coarse-grained ranking losses, respectively. The model is trained on samples across different stages and varying easy/hard sample levels, enabling it to achieve robust generalization during the pre-ranking stage while simultaneously optimizing hierarchical objectives.

Our synergistic framework is designed to learn both the hierarchy of user behaviors and the pattern of the cascade ranking system. For instance, in the context of online advertising, the model is expected to understand the following order of importance: converted items > clicked items > exposed items > candidate items and randomly sampled items. This prioritization helps ensure that the model effectively distinguishes between different levels of user engagement and optimizes the ranking accordingly.

\subsubsection{Distillation Loss}
As the main goal for the pre-ranking model is to output a high-quality item set for the ranking model, hence we used a listwise loss for distilling the knowledge from the ranking model as follows:

\begin{equation}
    \mathcal{L}_{\mathrm{Distillation}}(z, p) = -\sum_{i \in \mathcal{D}_I} p_i \log \frac{\exp(z_i)}{\sum_{j \in \mathcal{D}_I} \exp(z_j)} \\
\end{equation}

where $p$ is the prediction of the ranking model (soft label), $z$ is the logit of the pre-ranking model, $\mathcal{D_I}$ is the impression samples set.

It is important to note that in our approach, the distillation process from the ranking model to the pre-ranking model is conducted exclusively on impression samples. As the ranking model is trained solely on these impression samples, its ability to generalize to candidate samples and random samples is inherently limited. However, by carefully adjusting the weight of the distillation loss, we can enhance both the consistency between the models and the overall ranking capability of the pre-ranking model.

\subsubsection{Fine-Grained Ranking Loss}
The fine-grained ranking loss is applied to both impression samples and candidate samples. This loss is critically important during training as it directly corresponds to the sample scope used in serving. We employ the $\mathbf{SoftSort}$, a differentiable sorting loss, to learn user behavior and the patterns of the cascade ranking system. The primary objective of the fine-grained ranking loss is to precisely rank items according to the varying degrees of positive feedback they receive. Furthermore, this loss function is designed to effectively differentiate positives from impression samples and negatives from candidate samples.

Consider the $\mathbf{SoftSort}$ operator defined by metric function $\mathbf{d} = | \cdot |^p$ and temperature parameter $\tau$ for sorting $n$-dimensional real vectors $s \in \mathbb R^n$:

\begin{equation}
    \mathbf{SoftSort}^{d}_{\tau}(s) = \mathbf{softmax}(\frac{-\mathbf{d}(sort(s) \mathbf{1}^T, \mathbf{1}s^T)}{\tau})
\end{equation}

The output of SoftSort operator is a permutation matrix of dimension $n$. The softmax operator is applied row-wise, thereby relaxing the permutation matrices into a set of unimodal row-stochastic matrices. In simple words: \textit{the $r$-th row of the SoftSort operator is the \texttt{softmax} of the negative distances to the $r$-th largest element}. \cite{prillo2020softsort}

We then employ the softmax cross entropy between the permutation matrices of label $y$ and the permutation matrices of logit $z$. The $\mathbf{SoftSort}$ loss function is hereby defined as:

\begin{equation}
   \mathcal{L}_{\mathrm{Sorting}}(z, y) = - \mathbf{tr}\Big(\mathbf{J}_n
   \big(\mathbf{SoftSort}^{d}_{\tau}(y) \circ \log \mathbf{SoftSort}^{d}_{\tau}(z)\big)\Big)
\end{equation}

where $\mathbf{J}_n$ is a $n\times n$ matrix of ones, \(\mathbf{y} = (y_i)_{i \in \mathcal{D_I} \cup \mathcal{D_C}}\) is the hard label and \(\mathbf{z} = (z_i)_{i \in \mathcal{D_I} \cup \mathcal{D_C}}\) is the logit of the pre-ranking model. We use the $\mathbf{tr}$ to compute the element sum of the matrix $\mathbf{SoftSort}^{d}_{\tau}(y) \circ \log\big( \mathbf{SoftSort}^{d}_{\tau}(z) \big)$.

\subsubsection{Coarse-Grained Ranking Loss}
The coarse-grained ranking loss is applied to all the samples: impression samples, candidate samples and random negative samples. The primary objective of the coarse-grained ranking loss is to effectively separate positive samples from negative samples. Additionally, it supports the ranking process among positive samples by providing a framework for distinguishing varying degrees of relevance or quality within the positive sample set.

We propose the Adaptive Margin Rankmax (AM-Rankmax) as the coarse-grained ranking loss. The Adaptive Margin Rankmax Loss (AM-Rankmax Loss) is an innovative modification of the Rankmax loss function \cite{kong2020rankmax}, designed to enhance performance in ranking tasks. This loss function introduces an adaptive margin that varies based on the nature of the pair being compared and the label distance between the items in the pair, thereby extending the Rankmax loss to address ranking problems with ordered or continuous labels. This approach ensures better generalization and more accurate differentiation in ranking scenarios.

Consider the Rankmax loss for ranking problems with binary labels only:

\begin{equation}
\begin{aligned}
    \mathcal{L}_{Rankmax}(z, y) & = -\sum_{j:y_j>0}\log \text{Rankmax}(z, y)_j \\
    & = \sum_{j:y_j>0}\log \sum_{i=1}^{n} \big(z_i - z_j + 1\big)_{+}
\end{aligned}  
\end{equation}

The Rankmax loss is reminiscent of pairwise losses, the $z_i$ represents the predicted logit for the $i$-th item in the list, the $z_j$ is the predicted logit for the $j$-th target item. The objective of the Rankmax loss function is to ensure that the target item is ranked appropriately in relation to all other items in the list, including a margin term set to 1, which aids in enforcing the desired ranking separation.

To extend the Rankmax loss to more general ranking problems involving multi-level positive labels, we introduce the adaptive margin. The adaptive margin for Rankmax loss incorporates both the type of item pairs and the distance in their labels to enhance ranking performance:

\begin{itemize}
    \item The loss is applied only when $y_i < y_j$, which is more suitable for multi-level positive label scenario.
    \item The margin adjusts based on whether $y_i$ is positive or negative. When $y_i$ is negative, a larger margin is applied to ensure that the positive item associated $y_j$ is ranked significantly higher. Conversely, when $y_i$ is positive, a smaller margin is used to reflect subtle differences in relevance.
    \item The margin scales with the label distances between items. Greater distances in labels result in larger margins, ensuring proper ranking separation for items with significantly distinct labels.
\end{itemize}

The adaptive margin function 

\begin{equation}
\begin{aligned}
   m(i, j) = \alpha \cdot \mathbbm{I}(y_{i} = 0 ) + \delta(y_i, y_j)
\end{aligned}  
\end{equation}

where $\alpha$ is a constant for adding additional margin between negative and positive items, $\mathbbm{I}$ is the indicator function. The metric function $\delta$ can take various forms, for example $\delta(y_i, y_j) = 1$ or $\delta(y_i, y_j) = \beta |y_i - y_j|^p$. The adaptive margin Rankmax loss is then given by:

\begin{equation}
\begin{aligned}
    L_{\mathrm{AM-Rankmax}}(z, y) & = 
    \sum_{j: y_j > 0} \log \sum_{\substack{i : y_i < y_j}} \big(z_i - z_j + m(i, j) \big)_{+}
\end{aligned}  
\end{equation}

where \(\mathbf{y} = (y_i)_{i \in \mathcal{D_I} \cup \mathcal{D_C} \cup \mathcal{D_R}}\) is the hard label from all the samples and \(\mathbf{z} = (z_i)_{i \in \mathcal{D_I} \cup \mathcal{D_C} \cup \mathcal{D_R}}\) is the logit of the pre-ranking model.

By incorporating the adaptive margin function, the AM-Rankmax loss function can effectively adapt to scenarios with multiple positive labels of varying levels. This enhancement allows the model to handle different degrees of positive feedback, thereby improving its ability to generalize and accurately rank items in complex settings.

\subsubsection{The Hybrid Ranking Loss}
We design a hybrid ranking loss that integrates both distillation and ranking objectives. The hybrid ranking loss is the weighted sum of three losses:

\begin{equation}
\begin{aligned}
    \mathcal{L}_{\mathrm{Hybrid}}(z, y ) & = \lambda_1\mathcal{L}_{\mathrm{Distillation}}(z, p) \\
    & + \lambda_2\mathcal{L}_{\mathrm{Sorting}}(z, y) \\
    & + \lambda_3\mathcal{L}_{\mathrm{AM-Rankmax}}(z, y)
\end{aligned}
\end{equation}

where $\lambda_1$, $\lambda_2$ and $\lambda_3$ are the loss weights for each sub-objective. Balancing the distillation and ranking losses is crucial for the pre-ranking model to inherit the ranking model's capabilities while generalizing to broader sample spaces. Additionally, weighting the fine-grained and coarse-grained ranking losses appropriately ensures a balance between precise ranking of relevant items and overall retrieval robustness.

\section{Experiments}
In this section, we provide a comprehensive description of our experiments, including detailed information about the datasets, evaluation metrics, comparisons with state-of-the-art pre-ranking models, and corresponding analyses. The experiment results, conducted on three publicly available large-scale datasets spanning the domains of online advertising, e-commerce, and short video recommendation, demonstrate the effectiveness of RankTower in the domain of pre-ranking.

During experiments, we focus on evaluating the effectiveness of our proposed models and answering the following questions:

\begin{itemize}
    \item \textbf{Q1}: How does our proposed RankTower perform for pre-ranking task? Is it effective and efficient under extremely high-dimensional and sparse data settings?
    \item \textbf{Q2}: How do different settings on dataset sampling and training losses influence the performance of RankTower?
\end{itemize}

\subsection{Experiment Setup}

\subsubsection{Datasets}
We evaluate our proposed model using three publicly available real-world datasets commonly utilized in research: Alimama, Taobao and KuaiRand. For each dataset, we retain only users who have had at least 100 impressions and 20 instances of positive feedback. The data is randomly divided into three subsets: 70\% for training, 10\% for validation, and 20\% for testing. Since all labels in the aforementioned dataset are binary, we simply aggregate them by summing the labels to form the hard label.

\begin{itemize}
    \item \textbf{Alimama\footnote{https://tianchi.aliyun.com/dataset/408}} is a Alimama advertising dataset, which are displayed on the website of Taobao.
    \item \textbf{Taobao\footnote{https://tianchi.aliyun.com/dataset/649}} is a Taobao E-commerce dataset released Alibaba.
    \item \textbf{KuaiRand\footnote{https://kuairand.com/}} is a recommendation dataset collected from the video-sharing mobile app Kuaishou.
\end{itemize}

\subsubsection{Evaluation Metrics}
We consider Recall@K and NDCG@K for evaluating the performance of the models, and we set $k$ to 100 for all experiment metrics.

\textbf{Recall@K} is the fraction of relevant retrieved within the top $K$ recommendations. It's mainly used for measuring ranking system's capability on retrieving relevant items.

\textbf{NDCG@K} measures the quality of the ranking by considering both the relevance and the position of items within the top $K$ recommendations. Items with higher relevance ranked at higher position contribute more to the metric.

\subsubsection{Competing Models}
We compare RankTower with the following pre-ranking models: LR \cite{mcmahan2013ad}, Two-Tower \cite{huang2013learning}, DAT \cite{yu2021dual}, COLD \cite{wang2020cold}, IntTower \cite{li2022inttower} and ARF\cite{wang2023adaptive}.

\subsubsection{Reproducibility}
We implement all the models using Tensorflow~\cite{abadi2016tensorflow}. The mini-batch size is 64, and the embedding dimension is $\max(\lfloor \log_2(\text{cardinality})\rfloor, 16)$ for all the features. For optimization, we employ Adam~\cite{kingma2014adam} with a learning rate set to 0.001 for all the neural network models, and we apply FTRL~\cite{mcmahan2011follow, mcmahan2013ad} with a learning rate of 0.01 for LR. Grid-search for each competing model's hyper-parameters is conducted on the validation dataset. The number of DNN layers is from 1 to 4. The number of neurons ranges from 64 to 512. All the models are trained with early stopping and are evaluated every 1000 training steps.

For the hyper-parameters search of RankTower, The number of layers in a Multi-Head Gated Network is from 1 to 4. For the number of sub-spaces $H_{u}$ and $H_{i}$, the searched values are from 2 to 10. The number of units for all dense layers is from 32 to 256. Fot the metric function $d$ of the differentiable sorting loss, we use the squared distance function. For the $\alpha$ and metric function $\delta$ of in AM-Rankmax loss, we search for $\alpha$ in the range from 2 to 5, and set $\delta(y_i, y_j) = 1$ for simplicity.

For generating candidate samples, we first construct a two-tower model based on the approach described by \cite{covington2016deep}. We then utilize this model to retrieve candidates for each user, filtering out the items they have interacted with to build candidate samples. These candidate samples serve as hard negative cases for each user in our experiments.

For generating ranking model's prediction on impression samples, we ensemble several state-of-the-art ranking models \cite{guo2017deepfm,shan2016deep,yan2020xdeepint,wang2021dcn}. We use weighted logloss \cite{covington2016deep} as the training objective and apply a weighted average to ensemble all the models. All the ranking models are trained on the training dataset, and the ensemble weights are optimized on the validation dataset.

\subsection{Model Performance Comparison (Q1)}

\begin{table}[H]
    \caption{Performance Comparison of Different Algorithms on Alimama , Taobao and KuaiRand Dataset.}
    \label{tbl_model_performance}
    \centering
    \resizebox{1.0\linewidth}{!}{
        \begin{tabular}{ccccccc}
            \hline
            & \multicolumn{2}{c}{Alimama} & \multicolumn{2}{c}{Taobao} & \multicolumn{2}{c}{KuaiRand} \\
            Model & Recall@K & NDCG@K & Recall@K & NDCG@K & Recall@K & NDCG@K \\
            \hline
            LR & 0.4802 & 0.3237 & 0.4792 & 0.2685 & 0.6713 & 0.5027 \\
            Two-Tower & 0.5123 & 0.3428 & 0.5019 & 0.2921 & 0.6902 & 0.5258 \\
            DAT & 0.5161 & 0.3472 & 0.5089 & 0.3013 & 0.6955 & 0.5312 \\
            COLD & 0.5210 & 0.3518 & 0.5123 & 0.3070 & 0.7011 & 0.5349 \\
            IntTower & 0.5215 & 0.3519 & 0.5101 & 0.3051 & 0.6960 & 0.5309 \\
            ARF & 0.5318 & 0.3655 & 0.5215 & 0.3117 & 0.7096 & 0.5497 \\
            \hline
            RankTower & \textbf{0.5462} & \textbf{0.3794} & \textbf{0.5301} & \textbf{0.3223} & \textbf{0.7182} & \textbf{0.5551} \\
            \hline
        \end{tabular}
    }
\end{table}

The overall performance of different model architectures is listed in \Cref{tbl_model_performance}. We have the following observations in terms of model effectiveness:
\begin{itemize}[leftmargin=10pt]
    \item LR exhibits the lowest performance compared to the other neural network-based models.
    \item Two-Tower brings the most significant relative improvement in performance with increased model complexity relative to the LR baseline, highlighting the importance of learning deep feature interactions.
    \item While most of the baseline models are of two-tower architecture, COLD achieves relatively strong performance among the competing models, indicating the significance of learning user-item feature interactions.
    \item ARF outperform other models that do not utilize listwise ranking losses, highlighting the importance of listwise ranking losses for pre-ranking models.
    \item RankTower achieves the best prediction performance among all models. Our model's superior performance could be attributed to RankTower's effectively modeling of bi-directional user-item feature interactions, as well as the design of full-stage sampling and hybrid loss functions.
\end{itemize}

\subsection{Model Study (Q2)}
In order to have deeper insights into the proposed model, we conduct experiments on the KuaiRand dataset and compare model performance on different settings. This section evaluates the model performance change with respect to settings that include: 1) the effect of full-stage data sampling; 2) the effect of listwise ranking losses; 3) the effect of distillation from ranking model;

\subsubsection{Effect of Full-Stage Sampling}
To better understand the contribution of each component of the full-stage sampling strategy, we conduct a comprehensive ablation study. This study aims to isolate and evaluate the impact of each sampling component on the overall performance of the model. By systematically removing each component, we can identify its significance and contribution to the model's effectiveness.

As illustrated in \autoref{tbl_sampling}, the full-stage sampling strategy achieves the best overall performance. When the pre-ranking model is trained solely with impression samples, it fails to generalize to unexposed items, adversely affecting retrieval performance. Furthermore, we observe that candidate samples are more important than random samples. As hard negatives, candidate samples significantly enhance the model's ability to discriminate between relevant and non-relevant items.

\begin{table}[H]
    \caption{Experiment Results for Different Sampling Strategies.}
    \label{tbl_sampling}
    \centering
    \begin{tabular}{l|cc}
        \hline
                         & Recall@K & NDCG@K \\
        \hline
        Full-Stage Sampling    & 0.7182   & 0.5551 \\
        w/o random samples   & 0.7125   & 0.5437 \\
        w/o candidate samples   & 0.7040   & 0.5401 \\
        w/o candidate \& random samples   & 0.6981   & 0.5323 \\
        \hline
    \end{tabular}
\end{table}

\subsubsection{Effect of Listwise Ranking Losses}
In order to better understand the properties of the proposed hybrid loss, we compare the proposed hybrid loss with several widely used ranking losses in the industry. The experiment results, as shown in Table \ref{tbl_loss}, indicate that the hybrid loss consistently outperforms other alternatives. Notably, it surpasses both its individual components: the Sorting loss and the AM-Rankmax loss. Additionally, our proposed AM-Rankmax demonstrates superior performance compared to the original Rankmax loss and the Softmax loss.

\begin{table}[H]
    \caption{Experiment Results for Different Ranking Losses.}
    \label{tbl_loss}
    \centering
    \begin{tabular}{l|cc}
        \hline
                        & Recall@K & NDCG@K \\
        \hline
        Hybrid Loss    & 0.7182   & 0.5551 \\
        Sorting   & 0.7128   & 0.5516 \\
        AM-Rankmax   & 0.7132   & 0.5507 \\
        Rankmax   & 0.7105   & 0.5492 \\
        Softmax   & 0.7109   & 0.5498 \\
        ApproxNDCG   & 0.7006   & 0.5436 \\
        RankNet   & 0.7072   & 0.5452 \\
        \hline
    \end{tabular}
\end{table}

\subsubsection{Effect of Distillation from Ranking Model}
For improving the performance and training efficiency of the pre-ranking model, we conduct knowledge distillation from the raking model, utilizing the Softmax loss. Here we are doing ablation study on the distillation component and further compare Softmax loss with other alternatives.

The \autoref{tbl_distillation} demonstrate the efficacy of the ranking model in transferring knowledge to the pre-ranking model through the distillation process on the impression samples. Among various loss function experimented for distillation, the Softmax loss outperforms the other alternative losses. The Softmax loss, being a listwise ranking loss, proved more adept at distilling the ranking model's capabilities compared to the weighted logloss, which essentially is a pointwise approach and exhibited suboptimal performance in learning the relative ranking distribution. In contrast, the pairwise logloss, focusing solely on pairwise ordering of ranking model's predictions without considering the relative proximity of predictions, exhibited overfitting to the ranking model's outputs.

\begin{table}[H]
    \caption{Experiment Results for Different Distillation Losses.}
    \label{tbl_distillation}
    \centering
    \begin{tabular}{l|cc}
        \hline
                         & Recall@K & NDCG@K \\
        \hline
        Distillation (Softmax)    & 0.7182   & 0.5551 \\
        Distillation (Weighted Logloss)    & 0.7130   & 0.5519 \\
        Distillation (Pairwise Logloss)    & 0.7071   & 0.5432 \\
        No Distillation   & 0.7108   & 0.5495 \\
        \hline
    \end{tabular}
\end{table}

\section{Conclusion}
This paper introduces the RankTower model, designed to enhance the performance of the two-tower model by effectively capturing latent interactions between user and item. The RankTower architecture consists of a Multi-Head Gated Network and a Gated Cross-Attention Network, which model diversified latent user-item representations and captures complex user-item interactions dynamically. Additionally, the Maximum Similarity Layer contributes to improved serving efficiency. To ensure consistency with existing casecade ranking system, a hybrid loss function and full-stage sampling approach are integrated into the model's optimization framework. Comprehensive experiments demonstrate that RankTower significantly outperforms state-of-the-art pre-ranking models. In future work, we aim to study how to effectively and jointly optimize the cascade ranking system in an end-to-end fashion.

\bibliographystyle{ACM-Reference-Format}
\bibliography{ranktower.bib}

\end{document}